\documentclass[11pt]{article}
\pdfoutput=1
\usepackage{jheppub}
\usepackage{amsmath}
\usepackage{bm}
\usepackage{slashed}
\usepackage{graphicx}

\newlength{\dummysp}
\newcommand{\beq}{\begin{eqnarray}}
\newcommand{\eeq}{\end{eqnarray}}

\newcommand{\gappeq}{\mathrel{\rlap {\raise.5ex\hbox{$>$}}
{\lower.5ex\hbox{$\sim$}}}}
\newcommand{\lappeq}{\mathrel{\rlap{\raise.5ex\hbox{$<$}}
{\lower.5ex\hbox{$\sim$}}}}

\newcommand{\ben}{\begin{enumerate}}
\newcommand{\een}{\end{enumerate}}

\newcommand{\bit}{\begin{itemize}}
\newcommand{\eit}{\end{itemize}}

\def\[{\left [}
\def\]{\right ]}
\def\({\left (}
\def\){\right )}
\def\R{{\mathbb R}}
\def\S{{\mathbb S}}
\def\Z{{\mathbb Z}}

\usepackage{hyperref}

\title{Deconfinement on axion domain walls }
   
 \author[a]{Mohamed M. Anber,}\author[b]{Erich Poppitz} 
  
\affiliation[a]{Department of Physics, Lewis $\&$ Clark College, Portland, OR 97219, USA}
\affiliation[b]{Department of Physics,   University of Toronto, 
Toronto, ON M5S 1A7, Canada}
\emailAdd{manber@lclark.edu}\emailAdd{poppitz@physics.utoronto.ca}    
\abstract{ 

{\flushleft{W}}e study $SU(N_c)$ gauge theories with Dirac fermions in representations ${\cal{R}}$ of nonzero $N$-ality, coupled to axions. These theories have an exact discrete chiral symmetry, which has a mixed 't Hooft anomaly with general baryon-color-flavor backgrounds, called the ``BCF anomaly" in \cite{Anber:2019nze}.  The infrared theory also has an emergent $\mathbb Z_{N_c}^{(1)}$  $1$-form center symmetry. We show that the BCF anomaly is matched in the infrared    by  axion domain walls. We argue that $\mathbb Z_{N_c}^{(1)}$ is spontaneously broken on  axion domain walls, so that  nonzero $N$-ality Wilson loops obey the perimeter law and  probe quarks  are deconfined on the walls.  We give further support to our conclusion by using a calculable small-circle compactification to study the multi-scale structure of the axion domain walls and the microscopic physics of  deconfinement  on  their worldvolume. 
 
}

\begin{document}

\maketitle

\flushbottom

\section{Introduction }

{\flushleft{\bf Motivation:}} Understanding of the phase structure of gauge theories is an interesting and unsolved problem. Due to strong coupling, determining the renormalization group (RG) flow to the infrared (IR) is difficult and few general constraints on  possible IR behaviors exist. The 't Hooft anomaly matching \cite{tHooft:1979rat} stands out as  an exact constraint on the IR behavior:  the anomaly is RG invariant and any proposed IR phase has to produce the same anomalies as those of the ultraviolet (UV) theory. The idea of anomaly matching is not new and it has been an important tool constraining the behavior of strongly coupled  theories. Anomaly matching has recently attracted renewed interest due to the injection of substantial new insight \cite{Gaiotto:2014kfa,Gaiotto:2017yup,Gaiotto:2017tne}: that introducing background fields for all global symmetries---internal or spacetime, continuous or discrete, $0$- or higher-form---consistent with their faithful action, can yield new  nontrivial constraints on the possible IR phases. 
We  refer to these constraints as ``generalized 't Hooft anomalies." Their study is  evolving too rapidly to allow us to do justice to all interesting aspects currently investigated; we only note that  complementary aspects of theories closely related to the ones we consider here are the subject of \cite{Komargodski:2017smk,Tanizaki:2018wtg,Wan:2018bns, Cordova:2019jnf, Cordova:2019uob,Anber:2019nfu,Bolognesi:2019fej,Cordova:2019bsd,Hason:2019akw,Wang:2019obe,Wan:2019soo,Wan:2019oax,Cordova:2019jqi}.

In this paper, we continue our study  of generalized 't Hooft anomalies in $SU(N_c)$  gauge theories with vectorlike fermions in representations of nonzero $N$-ality. These theories have exact discrete chiral symmetries. Earlier   \cite{Anber:2019nze}, we showed that introducing general baryon, flavor, and color (BCF) backgrounds, consistent with the faithful action of the global symmetries, leads to a mixed anomaly between the discrete chiral symmetry and the BCF background. We used this ``BCF anomaly" to rule out certain kinds of IR behavior---notably, we showed that some gauge theories cannot flow to an IR theory of massless composite fermions only. 

{\flushleft{\bf Summary of results:}} Here, we consider  the same class of theories in  cases when the  discrete chiral symmetry is broken.  Concretely, we focus on the breaking of the chiral symmetry and the matching of the BCF anomaly after coupling the theory to an axion. We take the  axion scale $v$ larger than the strong coupling scale  $\Lambda$ of the theory, so that the IR theory is that of only a light axion, of mass $\Lambda^2 \over v$. We argue that the BCF anomaly is matched by axion domain walls that arise from the breakdown of the chiral symmetry. Anomaly matching also requires nontrivial physics on their worldvolume: Wilson loops are expected to obey perimeter law on the domain walls, corresponding to the breaking of an (emergent) $1$-form center symmetry and the deconfinement of quarks on the domain wall worldvolume.

Elucidating the microscopic nature of the worldvolume physics is the main goal of this paper. This is not  straightforward  on $\R^4$, as the physics determining the domain wall behavior is not semiclassically accessible, contrary to naive expectations---despite the axion's lightness, axion domain walls probe also the much  shorter distance scales of order the confinement scale, representing a subtle failure of effective field theory (EFT) that has been anticipated earlier. In particular,  it was argued, using large-$N_c$ arguments  (see  \cite{Gabadadze:2002ff} for a review and \cite{Komargodski:2018odf} for recent related work) that as the axion wall is traversed, a rearrangement of the hadronic degrees of freedom should take place. 

We show that the deconfinement and  rearrangement of heavy degrees of freedom on the axion domain wall are intertwined. 
We use a calculable compactification \cite{Unsal:2007jx,Unsal:2008ch} to $\R^3 \times \S^1$ to study the mechanism of deconfinement, following the  observations of our work with Sulejmanpasic \cite{Anber:2015kea} and the recent remarks  on anomaly matching \cite{Tanizaki:2019rbk}. Here, the physics  of confinement is semiclassical and the failure of the axion EFT can be traced to the multi-branch nature of its potential (which, in this setup, is seen at any finite $N_c$). 
We show that elementary axion domain walls necessarily involve  ``branch hopping," and as a result they acquire a multi-scale structure, probing both the axion scale and the much shorter confinement scale. This ``layered" structure of the axion domain wall is ultimately responsible for the nontrivial worldvolume physics leading to quark deconfinement, which is otherwise similar to \cite{Anber:2015kea,Cox:2019aji}. 

Our results give an explicit realization of domain wall anomaly inflow,  yield further confidence both in the generalized anomaly arguments and the ``adiabatic continuity"  of  $\R^3\times \S^1$ compactifications  \cite{Dunne:2016nmc}, and suggest  that multi-scale structures on axion domain walls also appear on $\R^4$ at finite $N_c$.

{\flushleft{\bf Organization of this paper:}} In Section \ref{bcfsection}, we describe the class of theories we study, their coupling to the axion, the symmetries, and the expected axion dynamics on $\R^4$. We also review the BCF anomaly, show that it is matched in the IR by axion domain walls, and discuss the nontrivial   physics and deconfinement  on their worldvolume. In Section \ref{calculabledeformation}, we study the same theories on $\R^3 \times \S^1$, with $\S^1$ size $L$ obeying $L N_c \Lambda \ll 1$, a condition making semiclassical analysis  possible. We review the salient features of such compactifications and use them to study the BCF anomaly matching, the vacuum structure, and domain walls of the axion theory. We then explain the multi-scale structure on axion domain walls and the physics responsible for the deconfinement of quarks on the domain walls.

{\flushleft{\bf In lieu of conclusion:}}
Many interesting questions remain unanswered. For example, a possible phase of these theories, with the fermions  taken  light or massless, is also one of broken chiral symmetry. Thus, it would  be interesting to study the behaviour of the theory, in the bulk and on the domain walls, as the axion scale is lowered, see \cite{Choi:2018tuh} for related work. The finite temperature phases are also of interest.  Our results here also lead us to expect that physics on walls between non-neighboring vacua may have interesting properties. Further, formulating  these theories on more general spacetime manifolds than the ones considered here may also  lead to more stringent generalized 't Hooft anomalies. Finally, various aspects of the physics discussed may be relevant for  ``hidden sector" models of dark matter and models of natural inflation. We hope to return to these questions in the future.

\section{The BCF anomaly with axions}
\label{bcfsection}

{\flushleft{\bf Theories and symmetries:}} In this paper, we consider $SU(N_c)$ gauge theories with $N_f$ flavors of vectorlike fermions $(\psi, \tilde\psi)$, both taken as two-component left-handed Weyl fermions, transforming in the representation $(\cal{R}, \cal{\overline{R}})$ of $SU(N_c)$ and the (anti) fundamental $(\square, \overline\square)$  representation of the vectorlike $U(N_f) = {SU(N_f) \times U(1)_B \over \Z_{N_f}}$ flavor symmetry. We take the $U(1)_B$ charges of $(\psi, \tilde\psi)$ to be $(+1, -1)$.  We denote by $n_c$ the $N$-ality of the representation $\cal{R}$, the number of boxes in the Young tableau of $\cal{R}$ modulo $N_c$. We focus   on representations of nonzero $N$-ality. 

The fermions are further coupled to a complex Higgs field $\Phi = \rho e^{i a} = \phi_1 + i \phi_2$, neutral under the $SU(N_c)$ gauge group. The Higgs field has potential $V(\Phi) = \lambda (|\Phi|^2 - v^2)^2$ and a Yukawa coupling  to the fermions,\footnote{In terms of $N_f$ 4-component  Dirac  fermions $\Psi$   in ${\cal{R}}$ of $SU(N_c)$, the Yukawa coupling is $y   \bar\Psi (\phi_1 + i \phi_2 \gamma_5)  \Psi$.} 
$L_{Y} = y \Phi   \tilde\psi \cdot \psi  + {\rm h.c.}$ The Yukawa coupling respects both the $U(N_f)$ global symmetry and the classical $U(1)$ global chiral symmetry: $\Phi \rightarrow e^{2 i \alpha} \Phi$, $\psi \rightarrow e^{-   i \alpha} \psi$, $\tilde\psi \rightarrow e^{- i \alpha} \tilde\psi$. The $U(1)$ chiral symmetry is broken by the anomaly to $\Z_{2 N_f T_{\cal{R}}}$, leaving only invariance under
\begin{equation}
\label{globalchiral}
\Z_{2 N_f T_{\cal{R}}}: ~ \Phi \rightarrow e^{i   {4 \pi k \over 2 N_f T_{\cal{R}}} }~ \Phi,~~ \psi \rightarrow e^{- i {2 \pi k \over 2 N_f T_{\cal{R}}} } ~\psi,~~ \tilde\psi \rightarrow  e^{- i {2 \pi k \over 2 N_f T_{\cal{R}}} } ~\tilde\psi~.
\end{equation}
The Dynkin index $T_{\cal{R}}$ of the representation $\cal{R}$ is normalized so that for the fundamental representation   $T_{\square} = 1$. In addition, when gcd$(N_c,n_c) = p>1$ these theories have an exact $\Z_p^{(1)}$ 1-form center symmetry.

It is useful to keep in mind a few simple benchmark cases: {\bf i.)}  when $\cal{R}$ is   the fundamental (F) representation, the theory we study is $N_f$-flavor $SU(N_c)$ QCD(F) coupled to the scalar $\Phi$, with a $\Z_{2 N_f}$ anomaly free global chiral symmetry (\ref{globalchiral}).\footnote{For $N_f=1$  QCD(F), only  fermion number $\Z_2$ is anomaly free; the Yukawa coupling and a fermion mass term then have the same symmetries.}  and {\bf ii.)} when $\cal{R}$ is the two-index symmetric (S) or antisymmetric (AS) representation, the theory  is  $N_f$-flavor $SU(N_c)$ QCD(S/AS), coupled to $\Phi$, with a $\Z_{2 N_f (N + 2)}$ (S) or $\Z_{2 N_f (N - 2)}$ (AS) anomaly free symmetry (\ref{globalchiral}). For $N_c$-even, the theories of type {\bf ii.} have a $1$-form $\Z_2^{(1)}$ center symmetry, reflecting the fact that two-index representation fields can not screen fundamental  quarks.

As stated in the Introduction, we shall be concerned with the regime $v \gg \Lambda, y \sim {\cal{O}}(1)$, with $\Lambda$  the strong coupling scale of the gauge theory. The Higgs field, of charge 2 under $\Z_{2 N_f T_{\cal{R}}}$ (as per (\ref{globalchiral})), acquires an expectation value $\langle \Phi \rangle = v$, breaking the global symmetry $\Z_{2 N_f T_{\cal{R}}} \rightarrow \Z_2$, where $\Z_2$ is the fermion number. At long distance, only the axion $a$ survives (as $\Phi \approx v e^{i a}$).  In the regime of interest, the fermions can be integrated out, giving rise to an effective theory at scales $\Lambda \ll \mu \ll v$:
\begin{equation}\label{effective1}
L_{\Lambda \ll \mu \ll v} = {v^2 \over 2} (\partial_\mu a)^2 +   N_f T_{\cal{R}} \; a \; q^c + L_{gauge}~+ \ldots~.
\end{equation}
Here we denoted by $q^c$ and $L_{gauge}$ the topological charge density\footnote{The topological charge is $Q^c = \int d^4 x \;q^c  \in \Z$ for the dynamical fields of the $SU(N_c)$ gauge theory.} and kinetic term of the $SU(N_c)$ gauge field, respectively, and the dots denote higher dimensional operators suppressed by the Higgs and fermion mass. In the absence of an anomaly, the axion $a$ would be a $2\pi$-periodic Goldstone field of the spontaneously broken $U(1)$ symmetry. The coupling to the topological charge density $q^c$ breaks the axion shift symmetry  to the discrete subgroup
(\ref{globalchiral}) 
\begin{equation}\label{axionshift}
a \rightarrow a + {2 \pi  \over   N_f T_{\cal{R}}}~.
\end{equation}
At scales $\mu \le \Lambda$, we can also integrate out the gauge field fluctuations, and one expects an effective Lagrangian for the axion field only:
\begin{equation}\label{effective2}
L_{ \mu \ll \Lambda} = {v^2 \over 2} (\partial_\mu a)^2 + \Lambda^4\;(1- \cos( a \;  N_f T_{\cal{R}})) + \ldots~,
\end{equation}
where  dots denote other terms in the (non-calculable) periodic axion potential. The theory (\ref{effective2}) has $N_f T_{\cal{R}}$ vacua corresponding to the $\Z_{2 N_f T_{\cal{R}}} \rightarrow \Z_2$ symmetry breaking. The vacua are gapped, with the axion mass of order $m_a \sim {\Lambda^2 N_f T_{\cal{R}}\over v}$. The potential in  (\ref{effective2}) is---at best, see below---only a model for the long-distance dynamics. Nonetheless, it is natural to expect  that any potential with the proper periodicity will exhibit this symmetry breaking pattern, giving rise to $N_f T_{\cal{R}}$ vacua and the associated domain walls (DW).   Taking (\ref{effective2}) at face value, one infers that the DW width  scales as $\delta_{DW}   \sim {1\over \Lambda} {v \over \Lambda N_f T_{\cal{R}}}\gg {1\over \Lambda}$. 

Thus, in the regime we study,  the  naive  axion effective action (\ref{effective2})   indicates that the DW dynamics involves distance scales larger than the inverse  strong-coupling scale $\Lambda$. However, as we shall see, the naive action (\ref{effective2}) does not account for important physical phenomena detected by long-distance probes. In particular, anomaly considerations will lead us to expect that the DW structure probes also distance scales of order $\Lambda^{-1}$, contrary to the naive expectation from (\ref{effective2}).\footnote{The expectation that axion DW have a more complicated structure than implied by a naive deduction from (\ref{effective2}) is not new---see  \cite{Gabadadze:2002ff} for a review of   QCD with light quarks, large-$N_c$ limit, super-Yang-Mills, and $D$-branes, and \cite{Komargodski:2018odf} for a related recent  study. The new elements in our  discussion are  the use of generalized 't Hooft anomalies and the explicit demonstration, see Section \ref{calculabledeformation}, of the DW structure and the deconfinement of quarks.} As we shall see in  Section \ref{calculabledeformation}, this expectation is  explicitly confirmed   in a calculable deformation of the axion theory.

Postponing momentarily a discussion of these topics, we  continue by describing the dynamics at longer length scales. At distances longer than the axion Compton wavelength $m_a^{-1}$ (formally, in the $v  \rightarrow \infty$ limit with fixed ${v\over \Lambda} \gg 1$) the $N_f T_{\cal{R}}$ vacua of (\ref{effective2})
 can be described by means of a $\Z_{N_f T_{\cal{R}}}$ TQFT (see recent discussion in \cite{Hidaka:2019mfm}). The fields in the TQFT are a compact scalar $\phi^{(0)}$ and a compact 3-form field $a^{(3)}$ (such that $\oint d \phi^{(0)}\in 2 \pi \Z$, $\oint da^{(3)}\in 2 \pi \Z$, where the integrals are over closed manifolds of the appropriate dimensionality) with Minkowskian Lagrangian
\begin{equation}
\label{bulkTQFT}
L_{TQFT} = {N_f T_{\cal{R}} \over 2 \pi}\; \phi^{(0)} \wedge d a^{(3)}~.
\end{equation}
 The TQFT (\ref{bulkTQFT}) has a $0$-form $\Z_{N_f T_{\cal{R}}}$ global symmetry $\phi^{(0)} \rightarrow \phi^{(0)} + {2 \pi \over N_f T_{\cal{R}}}$ and a $3$-form $\Z_{N_f T_{\cal{R}}}^{(3)}$ global symmetry: $a^{(3)} \rightarrow a^{(3)} + {1 \over N_f T_{\cal{R}}}\epsilon^{(3)}$, where $\oint \epsilon^{(3)} \in 2 \pi \Z$ is a constant $3$-form with integral periods. There are $N_f T_{\cal{R}}$ vacua\footnote{This can be seen  upon canonical quantization, requiring fixing the gauge under the $2$-form gauge transformations of $a^{(3)}$ (see e.g.~\cite{Kapustin:2014gua} for details).} corresponding to the breaking of the $0$-form $\Z_{N_f T_{\cal{R}}}$ symmetry. The objects charged under the $3$-form symmetry are the DW between these vacua. We note that the TQFT carries no information on the energy   and length scales associated with the DW, which are determined by the UV theory, e.g.~(\ref{effective2}) or its generalization. Likewise,   the $3$-form symmetry, absent in (\ref{effective2}), is   emergent  at long distances (according to the general criteria of \cite{Gaiotto:2014kfa}, it is also broken---the operators $e^{i \oint a^{(3)}}$, with the integral taken over closed $3$-manifolds, obey a ``$3$-volume" law, the analogue of perimeter law for Wilson loops). The theory (\ref{bulkTQFT}) will be useful to study the IR matching of 't Hooft anomalies of the global symmetries.
 
{\flushleft{\bf Generalized  't Hooft anomalies:}} Recall that 
in \cite{Anber:2019nze}  we coupled the gauge theory considered here to background fields of the  global flavor symmetries. In fact, the backgrounds considered corresponded to gauging  the $U(N_f) \over  \Z_{N_c}$ global symmetry, where $\Z_{N_c}$ denotes the center of the gauge group. The modding by the center of the gauge group arises because some
discrete transformations of $U(1)_B$,  acting  on $\psi$ and $\tilde\psi$ with opposite phases,  are really part of the gauge group. Explicitly, when the theory is considered on the four-torus, these backgrounds are 't Hooft fluxes for the  $SU(N_c), SU(N_f)$, and $U(1)_B$ gauge fields. These 't Hooft fluxes are chosen to be consistent with the torus transition functions for the fermions ($\psi$, $\tilde\psi$) in the representation ($\cal{R}, \cal{\overline{R}}$).

 We shall not give the explicit form of the 't Hooft flux backgrounds here and refer the reader to \cite{Anber:2019nze}. The important point to make is that these $U(N_f) \over  \Z_{N_c}$ global symmetry\footnote{The part of the gauge group acting faithfully  on fermions of $N$-ality $n_c$ is $SU(N_c)/\Z_p$, with $p={\rm gcd}(N_c, n_c)$.  This means that the fermions are charged under $\Z_{N_c\over p}$ and the true global symmetry is    $U(N_f)/\Z_{N_c\over p}$ (where $U(N_f) = (SU(N_f) \times U(1)_B)/\Z_{N_f})$. The backgrounds with topological charges (\ref{backgrounds}) are the most general ones consistent with the transition functions of the $N$-ality $n_c$ fermions on the four-torus $\mathbb{T}^4$ \cite{Anber:2019nze}. } backgrounds carry fractional topological charges $Q^c$ of $SU(N_c)$, $Q^f$ of $SU(N_f)$, and $Q^B$ of $U(1)_B$:\begin{eqnarray}\label{backgrounds}
Q^c = m\;  m' \left( 1 - {1 \over N_c}\right),   ~ Q^f =  k\; k' \left(1 - {1 \over N_f}\right),~ Q^B = \left( n_c {m   \over N_c} + {k  \over N_f}\right) \left( n_c {m'   \over N_c} + {k'  \over N_f}\right), \end{eqnarray}
where $m, m'$ are integers defined modulo $N_c$ (likewise, $k, k' \in \Z$ are defined modulo $N_f$). The importance of the global symmetry backgrounds (\ref{backgrounds}) is that under the anomaly-free discrete chiral $\Z_{ 2 N_f T_{\cal{R}} }$ transformation (\ref{globalchiral}), the partition function $Z$ of the UV theory transforms as
\begin{eqnarray}
\label{BCF1}
 \Z_{2 N_f T_{\cal{R}}}: \;  Z  &\rightarrow& Z \; e^{i {2 \pi \over   N_f T_{\cal{R}} }\left[   N_f T_{\cal{R}} Q^c +    d_{\cal{R}}  Q^f +   N_f d_{\cal{R}} Q^B\right]}, \end{eqnarray}
 where $d_{\cal{R}}$ denotes the dimension of the representation $\cal{R}$.
It is easy to see that the phase on the r.h.s.~of (\ref{BCF1})  is nontrivial in the background (\ref{backgrounds}).\footnote{For a single flavor   of the two-index S  or AS representation, the phase in (\ref{BCF1}) is  $\mathbb Z_{N\pm2}$-valued.} Thus, the discrete chiral $\Z_{2N_f T_{\cal{R}}}$ symmetry   has a mixed 't Hooft anomaly with the $U(N_f) \over \Z_{N_c}$ symmetry. This mixed anomaly was termed the ``BCF anomaly" in \cite{Anber:2019nze}. The anomaly  was  shown to restrict the possible IR phases of these gauge theories.

Coming back to  the theory with axions, in the background  with topological charges (\ref{backgrounds}), the axion acquires also couplings to the topological charges of the various background fields:\begin{equation}\label{axiontop}
L_{top.} = a \; (N_f T_{\cal{R}} \; q^c +  d_{\cal{R}}  \; q^f +   N_f d_{\cal{R}} \;q^B),~
\end{equation}
where $q^f$ and $q^B$ are the topological charge densities of the flavor and baryon number fields and we included the coupling to $q^c$ from  (\ref{effective1}). The couplings of the axion to the  topological charge densities in (\ref{axiontop}) match  the BCF anomaly at the energy scales where (\ref{effective1}) is valid: under a $\Z_{N_f T_{\cal{R}}}$ shift of the axion (\ref{axionshift}), in the backgrounds (\ref{backgrounds}), the transformation (\ref{BCF1}) of the partition function is reproduced by (\ref{axiontop}).


The BCF anomaly in the bulk can be also given a description using the continuum formalism of \cite{Kapustin:2014gua}
describing the gauging of higher form symmetries. This  description can also be used at scales longer than the axion Compton wavelength, i.e.~applied to the TQFT (\ref{bulkTQFT}). We shall find it also useful in Section \ref{calculabledeformation} and so we briefly review it next.

The continuum formalism of gauging higher form symmetries uses an  embedding of the $SU(N_c)$ connection $A^c$  into  a $U(N_c)$ connection. 
We   review the construction  for $SU(N_c)$; it proceeds similarly for $SU(N_f)$. We take pairs of $U(1)$ $2$-form and $1$-form gauge fields $\left(B^{c(2)},B^{c(1)} \right)$ such that $dB^{c(1)}=N_cB^{c(2)}$. The $1$-form gauge fields satisfy $\oint dB^{c(1)} \in   2 \pi \Z$, where the integrals are taken over closed $2$ surfaces. Thus, we have $\oint B^{c(2)}\in \frac{2 \pi}{N_c}\mathbb Z$. Next, we define the $U(N_c)$ connections 
$
\tilde A^c\equiv A^c+\frac{B^{c(1)}}{N_c},
$
where the second term is proportional to the $N_c \times N_c$ unit matrix and $A^c$ is the $SU(N_c)$ connection.
The gauge field strengths $\tilde F^c=d\tilde A^c+\tilde A^c\wedge\tilde A^c$  satisfy $\mbox{tr}_F \tilde F^c =dB^{c(1)}=N_cB^{c(2)}$. Going from $SU(N_c)$ to $U(N_c)$ introduces extra degrees of freedom. In order to eliminate them, we postulate an invariance under $U(1)$ $1$-from gauge symmetries: $\tilde A^c\rightarrow \tilde A^c+\lambda^{c(1)}$, which translates into $\tilde F^c\rightarrow \tilde F^c+d\lambda^{c(1)}$. The   fields $\left(B^{c(2)},B^{c(1)} \right)$ transform as $B^{c(2)}\rightarrow B^{c(2)}+d\lambda^{c(1)}$, $B^{c(1)}\rightarrow B^{c(1)}+ N_c \lambda^{c(1)}$, so that the constraints $dB^{c(1)}=N_cB^{c(2)}$ are invariant (the $1$-form transformation parameter obeys $\oint d \lambda^{c(1)} \in 2 \pi \Z$).
Introducing  the background fields $(B^{c(1)}, B^{c(2)})$ into the Lagrangian of our gauge theory, as described above, is equivalent to turning on 't Hooft fluxes for $SU(N_c)$, with topological charge $Q^c$ from  (\ref{backgrounds}). 

Similarly, we introduce fields $(B^{f(1)}, B^{f(2)})$ for $SU(N_f)$,  along with the corresponding $1$-form symmetry with parameter $\lambda^{f(1)}$, coupled to the $SU(N_f)$ theory via the  $U(N_f)$ connection constructed as outlined above. The  $(B^{f(1)}, B^{f(2)})$ background  corresponds to the introduction of the 't Hooft flux for $SU(N_f)$, whose topological charge $Q^f$ is given in (\ref{backgrounds}).

Up to this stage, we have not yet said anything about the baryon background gauge field $A^B$. Invariance of the matter covariant derivatives of ($\psi, \tilde\psi$) under the $U(1)$ $1$-from gauge symmetries demands\footnote{This is equivalent to requiring that the transition functions for the matter fields in the 't Hooft flux backgrounds (\ref{backgrounds}) satisfy the cocycle conditions.} that $A^B\rightarrow A^B+n^c\lambda^{c(1)}+ \lambda^{f(1)}$. Using the field strength $F^B=dA^B$ we find $F^B\rightarrow F^B +n^c d\lambda^{c (1)}+ \lambda^{f(1)}$. 

Thus, we have the following $2$-form combinations 
$
\tilde F^c-B^{c(2)}$, $\tilde F^f - B^{f(2)}$, $F^B-n^cB^{c(2)}- B^{f(2)},
$
invariant under the $1$-form gauge transformations with parameters $\lambda^{c(1)}$ and $\lambda^{f(1)}$.
In terms of these, the topological charge densities appearing in (\ref{axiontop}) are\begin{eqnarray}
\nonumber
q^c &=&\frac{1}{8\pi^2}\left[ \mbox{tr}_F\left(\tilde F^c\wedge \tilde F^c\right)-N_c  B^{c(2)}\wedge B^{c(2)}  \right],~ q^f  =  \frac{1}{8\pi^2} \left[ \mbox{tr}_F\left(\tilde F^f\wedge \tilde F^f\right)-N_f B^{f(2)}\wedge B^{f(2)}  \right],  \\
q^B&=&\frac{1}{8\pi^2} \left[F^B-n^cB^{c(2)}-B^{f(2)} \right]\wedge \left[F^B-n^cB^{c(2)}- B^{f(2)} \right].\label{continuumQ} 
\end{eqnarray}
Integrating the above  $q^{c,f,B}$ over the four dimensional spacetime   gives rise  to the  topological charges
 in (\ref{backgrounds}). One notes that the $U(N_c)$, $U(N_f)$, and $U(1)_B$ topological charges (the terms proportional to $\int\mbox{tr}_F \tilde F^c\wedge \tilde F^c$, $\int\mbox{tr}_F \tilde F^f\wedge \tilde F^f$, $\int F_B \wedge F_B$, respectively) are integer  on spin manifolds, while the  terms containing $B^{c(2)}$ and $B^{f(2)}$ give rise to the fractional terms  in (\ref{backgrounds}), once the conditions $\oint B^{c(2)} = {2 \pi \Z \over N_{c}}$, and similar for $c \rightarrow f$, are taken into account.
 
 It is now clear that using (\ref{continuumQ}) in (\ref{axiontop}) reproduces the BCF anomaly (\ref{BCF1}) in the effective theory (\ref{effective1}). 
 Employing the above formalism, the TQFT action (\ref{bulkTQFT}) can also be coupled to the $U(N_f)\over \Z_N$ background  \begin{equation}
\label{bulkTQFT1}
L_{TQFT} = {N_f T_{\cal{R}} \over 2 \pi}\; \phi^{(0)} \wedge \left(d a^{(3)} -{N_c \over 4 \pi}  B^{c(2)}\wedge B^{c(2)} + 2 \pi {d_{\cal{R}} \over N_f  T_{\cal{R}}} \; q^f +2\pi  {d_{\cal{R}} \over  T_{\cal{R}}} \;q^B\right)~,
\end{equation}
and is easily seen to reproduce the BCF anomaly (\ref{BCF1}) upon a $\Z_{N_f T_{\cal{R}}}$ shift of  $\phi^{(0)}$ (recalling that $\oint da^{(3)}\in 2 \pi \Z$). Notice that making (\ref{bulkTQFT1}) invariant under the same $1$-form transformations that leave (\ref{axiontop}, \ref{continuumQ}) intact,   the $3$-form field  $a^{(3)}$ acquires a shift under $\Z_{N_c}^{(1)}$ center transformations  with parameters $\lambda^{c(1)}$ \cite{Gaiotto:2014kfa,Seiberg:2018ntt,Cordova:2019uob}:
\begin{equation}
\label{a3shift}
a^{(3)} \rightarrow a^{(3)} + {N_c \over 2 \pi} B^{c(2)} \wedge \lambda^{c(1)} + {N_c \over 4\pi} \lambda^{c(1)} \wedge d \lambda^{c(1)}.  
\end{equation}
Thus, with (\ref{a3shift}), the theory (\ref{bulkTQFT1}) is   invariant under the $1$-form transformations with parameters $\lambda^{c(1)}, \lambda^{f(1)}$ (acting on dynamical and background fields). As already argued, it also reproduces the BCF anomaly (\ref{BCF1}) upon a $\Z_{N_f T_{\cal{R}}}$ shift of $\phi^{(0)}$. 

{\flushleft{\bf Emergent center symmetry and deconfinement on DW:}}
In the limit we study,  $v \gg \Lambda$, the IR theories (\ref{effective1}, \ref{effective2}, \ref{bulkTQFT})  acquire an emergent $1$-form $\Z_{N_c}^{(1)}$ symmetry,\footnote{In addition to the possible $\Z_{p}^{(1)}$ center present in the UV theory when gcd($N_c, n_c) = p >1$.} as the fermions  (whose coupling to the gauge field is the only source breaking the center symmetry) decouple. In the case of noncompact $\R^4$, integrating out fermions generates local terms only.\footnote{In contrast, on $\R^3 \times \S^1$, operators winding the $\S^1$ can be important in the regime of interest.} On the other hand, there exist no local terms involving the light dynamical fields---the axion and the $SU(N_c)$ gauge fields---that violate the $\Z_{N_c}^{(1)}$ center symmetry, as the latter only acts on topologically nontrivial line operators. Thus, the long distance theory has an emergent   $\Z_{N_c}^{(1)}$ symmetry. In fact, this symmetry is manifest in the TQFT (\ref{bulkTQFT1}), where the transformation (\ref{a3shift}) is a symmetry (in the absence of the nondynamical baryon and flavor backgrounds). 

Similar to the case of DW between chirally broken vacua in super-Yang-Mills theory, or  DW between CP breaking vacua in $\theta=\pi$ Yang-Mills theory, one expects that this emergent $1$-form symmetry is also  broken on the axion DW,   hence quarks of all $N$-alities are deconfined. Formally, one  argues that  the DW between neighboring vacua carries an $SU(N_c)_1$ Chern-Simons theory which has a $\Z_{N_c}^{(1)}$ center symmetry with a 't Hooft anomaly, for details we refer the reader to  \cite{Gaiotto:2017yup,Gaiotto:2017tne,Hsin:2018vcg}. Heuristically, one can argue that this Chern-Simons theory arises because crossing the axion DW implements a $2\pi$ shift of the $\theta$ parameter \cite{Komargodski:2018odf}. 
Further, it is natural to identify Wilson loops in the worldvolume Chern-Simons theory with Wilson loops of the gauge theory, taken to lie in the DW worldvolume. Since 
Wilson loops in Chern-Simons theory are known to obey perimeter law, one concludes  that quarks are deconfined on the DW worldvolume.

The arguments in favor of deconfinement on axion DW have a somewhat formal flavor and it would be desirable to have a setup where they can be explained more physically. Naturally, this is difficult in nonsupersymmetric theories on $\R^4$, since confinement occurs at strong coupling and a tractable theory thereof is lacking.\footnote{Arguments employing  monopole-/dyon-condensation confinement (as in Seiberg-Witten theory) have been used to give a physical picture of quark deconfinement on DW (to the best of our knowledge, beginning with  \cite{SJRey:1998}, as cited in \cite{Witten:1997ep}). However,  in the nonsupersymmetric theories at hand these excitations are  ``somewhat elusive"---quoting  \cite{Witten:1997ep}, see \cite{Greensite:2011zz} for a review---hence these  arguments are heuristic at best. See also  \cite{Campos:1998db} for discussions using models of the IR super Yang-Mills dynamics. The beauty of the $\R^3 \times \S^1$ explanation given in Section \ref{calculabledeformation}  is that it only involves controllable semiclassical physics.}  
However, as we shall show in the next section, using a compactification of the theory on $\R^3\times \S^1$ (but keeping the locally four dimensional nature and all relevant symmetries intact!) the physical mechanism underlying deconfinement on axion DW can be made explicit. On the one hand, this gives us further confidence in arguments based on anomaly inflow,  and, on the other, it points towards a more complicated structure of axion DW on $\R^4$ than the one expected from the naive axion effective theory (\ref{effective2}).

In the next Section, we take this more explicit  approach and  discuss anomaly matching and the deconfinement of quarks on the axion DW, in a setup where the properties of the ground state of the theory and the associated DW can be studied semiclassically.

\section{The BCF anomaly and deconfinement on DW on $\R^3 \times \S^1$}
\label{calculabledeformation}

The calculable setup we discuss here is one where our theory is compactified on a small circle $\S^1$ of circumference $L$. We consider the limit where $v \gg {1\over N_c L} \gg \Lambda$. In addition, the theory is ``deformed" by adding massive adjoint fermions whose presence is needed to  ensure center stability in the $N_c L\Lambda \ll 1$ semiclassical limit.\footnote{To ensure center stability, one can also   add nonlocal ``double-trace deformations" along $\S^1$ to the compactified theory, but these can be seen to be  generated by massive adjoint fermions, with mass ${\cal O}(1/N_c L)$. We take the view that a dynamical explanation of the double-trace deformations is needed to ensure renormalizability of the theory.  The setup described above is known as ``deformed Yang-Mills theory" and we refer the reader to the original paper \cite{Unsal:2008ch} for details. See    \cite{Anber:2017pak} for interesting variations and a large list of references.
} We stress that  adding massive adjoints does not affect any of the axion couplings and discrete chiral symmetries discussed so far. 
 The effective potential for the $\S^1$-holonomy, to leading order in the $v \gg {1\over N_c L} \gg \Lambda$ limit, is  stabilized at the center symmetric value. The theory abelianizes, with  broken  $SU(N_c) \rightarrow U(1)^{N_c - 1}$ at a high scale $1\over N_c L$,  and  the long distance $\R^3$ physics  is  described in terms of the $N_c-1$ dual photons in the Cartan subalgebra of $SU(N_c)$. 

In the heavy-fermion theory on $\R^3 \times \S^1$, the $(\psi, \tilde\psi)$ fermions have mass $yv \gg 1/L$ and, as before, can be integrated out. When the theory is considered at distances $\gg N_c L$, it has a $0$-form center symmetry $\Z_{N_c}^{(0)}$, the ``component" of the emergent $1$-form center in $\R^4$ along the compact direction, as well as a $\Z_{N_c}^{(1)}$  $\R^3$ $1$-form center.  There are exponentially suppressed terms,  $\sim e^{- y v L}$, local in $\R^3$ but winding around $\S^1$, that break the $0$-form $\Z_{N_c}^{(0)}$ center symmetry,\footnote{Generically, they will lead to an exponentially small shift of the holonomy $\vec\phi$ away from the $Z_{N_c}^{(0)}$ center symmetric point $\vec\phi = 2 \pi \vec\rho/N_c$, which will then only respect the $Z_{p={\rm gcd}(N_c, n_c)}^{(0)}$ center symmetry. We ignore these effects here and leave their study for the future (in particular only Wilson loops which can not be screened by $N$-ality $n_c$ quarks will be deconfined on the wall). } which we ignore. As is the case on $\R^4$, no local terms breaking the $\Z_{N_c}^{(1)}$ $\R^3$ $1$-form emergent center symmetry appear.

Thus, at the center-stabilized  value of the holonomy, the theory is essentially deformed Yang-Mills theory
 coupled to an axion field\footnote{See \cite{Anber:2015bba} for an earlier study of axions coupled to deformed Yang-Mills theory.} $a$. The axion appears as a dynamical $\theta$ parameter and has a kinetic term given by the reduction of (\ref{effective2}) to $\R^3$, ${v^2 L \over 2} (\partial_\mu a)^2$. 
 To further discuss the long-distance dynamics, 
we now borrow the results of the recent study  of deformed Yang-Mills theory with $\theta$ parameter \cite{Tanizaki:2019rbk},  adding the dynamical axion field and its topological coupling to the background  fields from (\ref{axiontop}).

{\flushleft{\bf Kinetic terms and anomalies:}} The kinetic and topological  ($\tilde{q}^{f,B}$   denote appropriate $\R^3$ reductions of $q^{f,B}$ from (\ref{continuumQ}),  given below in (\ref{r3topological})) terms in the Euclidean $\R^3$ Lagrangian are:\begin{eqnarray}
\label{r3kinetic}
L_{kin.} &=& {v^2 L \over 2} |d  a|^2  + {1 \over 2 g^2 L}|d\vec{\phi} - N_c A^{c (1)} \vec{\nu}_1|^2 +{g^2 \over 8 \pi^2 L}\left|d \vec{\sigma} + {N_f T_{\cal{R}} a \over 2 \pi}(d \vec{\phi} - N_c A^{c(1)} \vec{\nu}_1)\right|^2 \nonumber \\
&&- i \; {N_c \over 2\pi} \;\vec{\nu}_1 \cdot d \vec{\sigma} \wedge B^{c (2)} + i a (d_{\cal{R}} \tilde{q}^f + N_f d_{\cal{R}} \tilde{q}^B)~.
\end{eqnarray}
Here, $L$ is the $\S^1$ circumference and $g$ is the $SU(N_c)$ gauge coupling, frozen at a scale of order $1\over N_c L$. 
The axion field is the one we already introduced and the scale $v$ is the one from the Higgs potential. 
 The arrows denote vectors in the Cartan subalgebra of $SU(N_c)$: $\vec{\sigma}$ represents the $N_c-1$ Cartan dual photons, which are  compact scalars taking values in the weight lattice of $SU(N_c)$ ($\vec\sigma \equiv \vec\sigma + 2 \pi \vec{w}_p$, where $\vec{w}_p$ denote the $N_c-1$ fundamental weight vectors) and $\vec{\phi}$ are the eigenvalues of the $\S^1$-holonomy. The latter is defined so that a fundamental Wilson loop along the $\S^1$ has eigenvalues\footnote{The careful reader might notice a slight difference in (\ref{r3kinetic}) from the formulae of \cite{Tanizaki:2019rbk}  stemming from the different form of the fundamental $\S^1$ Wilson loop we use; no aspects of the physics are changed.} $e^{- i \vec\phi \cdot \vec\nu_A}$, $A = 1, \ldots N_c$, where $\vec\nu_A$ are the weights of the fundamental representation, normalized so that $\vec\nu_A \cdot \vec\nu_B = \delta_{AB} - 1/N_c$. Under global $0$-form center symmetry transformations, the holonomy eigenvalues shift as $\vec\phi \rightarrow \vec\phi + 2 \pi \vec\nu_1$, transforming all Wilson loop eigenvalues by  a $\Z_{N_c}$ phase, as appropriate. Further, we have dimensionally reduced the four dimensional 2-form gauge field  
$B^{c(2), 4d} = A^{c(1)} \wedge {d x^3 \over L} + B^{c(2)}$ into a 3d $1$-form $A^{c(1)}$ and a 3d $2$-form $B^{c(2)}$ gauge fields.\footnote{For brevity and to reduce clutter, we do not explicitly indicate  the three dimensional nature of all forms that appear below. We also warn the reader against confusing the dynamical $4$d color field $A^{c}$ with $A^{c(1)}$, the 3d $1$-form part of $B^{c(2)}$.}

 The Lagrangian (\ref{r3kinetic}) is clearly invariant under the local  $0$-form center symmetry, acting as $\vec\phi \rightarrow \vec\phi + N_c \lambda^{(0)} \vec{\nu}_1$ and $A^{c(1)} \rightarrow A^{c(1)} + d \lambda^{(0)}$.  The global $0$-form center is recovered upon taking $\lambda^{(0)} = {2 \pi \over N_c}$. Invariance of $e^{- \int_{\R^3} L_{kin}}$ under $1$-form center transformations, $B^{c(2)} \rightarrow B^{c(2)} + d \lambda^{c(1)}$, is ensured by the fact that $\oint d \lambda^{c(1)} \in 2 \pi \Z$ and that the monodromies of the dual photon lie in the weight lattice, i.e.~$\oint d \vec{\sigma} \in 2 \pi \Z \vec{w}$, where $\vec{w}$ is any fundamental weight.

Most importantly for us, under a
 $\Z_{N_f T_{\cal{R}}}$ shift  (\ref{axionshift}) of the axion field, we have that
  $d \vec{\sigma} \rightarrow d \vec{\sigma} - d \vec{\phi} + N_c A^{c(1)} \vec{\nu}_1$, leading to $\delta L_{kin.} = i {N_c \over 2 \pi} \vec\nu_1 \cdot d \vec{\phi} \;B^{c(2)} - i{N_c \over 2 \pi} (N_c -1) A^{c(1)}\wedge B^{c(2)}$. The first term in the variation of $L_{kin.}$ gives no contribution to $e^{- \int_{\R^3} L_{kin}}$, since $\oint {d\vec{\phi}\over 2 \pi}$ lies in the root lattice \cite{Argyres:2012ka,Anber:2015wha} and $\oint N_c B^{c(2)} \in 2 \pi \Z$, while the last term, using  $\oint A^{c(1)} \in {2 \pi \Z \over N_c}$, $\oint B^{c(2)} \in {2 \pi \Z \over N_c}$, reproduces the color background contribution to the BCF anomaly (\ref{BCF1}).  
  
  Finally, the other (baryon and flavor) contributions to the BCF anomaly (\ref{BCF1}) are due to the last two terms in (\ref{r3kinetic}), where $\tilde{q}^f$ and $\tilde{q}^B$ are  $\R^3$-reductions of $q^f$ and $q^B$ from (\ref{continuumQ}):
  \begin{eqnarray}
  \tilde{q}^B &=&\frac{1}{8\pi^2} (d \phi^B - n_c A^{c (1)} - A^{f (1)}) \wedge (F^B - n_c B^{c (2)} - B^{f(2)}),~  \nonumber \\
  \tilde{q}^f &=& \frac{1}{8\pi^2}  (d \vec\phi^f - N_f A^{f (1)} \vec\nu_1^{(f)})\cdot \wedge (\vec{F}^f - N_f B^{f(2)} \vec\nu_1^{(f)}). \label{r3topological}
  \end{eqnarray}
  Here $\phi^B$ is the $\S^1$ holonomy of the $U(1)_B$ background (and $\oint d \phi^B \in 2 \pi \Z$)  and $\vec\phi^f$ are the Cartan components of the $\S^1$ holonomy of the $SU(N_f)$ background fields ($\oint d \vec{\phi}^f/2 \pi$ is in the root lattice of $SU(N_f)$). Likewise $F^B$ ($\oint F^B/2\pi \in \Z)$  is the $\R^3$ $2$-form field strength of $U(1)_B$ and $\vec{F}^f$ denotes the Cartan components of the $\R^3$ $2$-form field strength of $SU(N_f)$ ($\oint \vec{F}^f/2 \pi$ is in the root lattice of $SU(N_f)$). We note that to describe the BCF anomaly, it suffices to turn on Cartan components only; correspondingly, $\vec\nu_1^{(f)}$ denotes the weight of the fundamental of $SU(N_f)$. Invariance under the $1$-form transformations on $\R^4$ (with parameters $\lambda^{c(1)}, \lambda^{f(1)}$, described above (\ref{continuumQ})) reduces to invariance under $0$-form transformations on $\R^3$ with parameters $\lambda^{c (0)}, \lambda^{f (0)}$, under which: $\delta A^{c(1)} = d \lambda^{c(0)}$, $\delta A^{f(1)} = d \lambda^{f(0)}$, $\delta \phi^B  = n_c \lambda^{c(0)} + \lambda^{f(0)}$, $\delta \vec\phi^f = N_f \vec\nu_1^{(f)} \lambda^{f(0)}$ and $1$-form  transformations on $\R^3$ with parameters $\lambda^{c(1)}, \lambda^{f(1)}$, under which $\delta B^{c(2)} = d \lambda^{c(1)}$, $\delta B^{f(2)} = d \lambda^{f(1)}$, $\delta \vec{F}^f = N_f d \lambda^{f(1)} \vec{\nu}_1^{(f)}$, $\delta F_B = n_c d \lambda^{c(1)} + d \lambda^{f(1)}$. It is clear that with the above transformations and the definitions (\ref{r3topological}), the $\R^3$ effective lagrangian (\ref{r3kinetic}) is invariant under the $\Z_{N_c}$ and $\Z_{N_f}$ transformations ($0$-form and $1$-form on $\R^3$) acting on the dynamical and background fields, as is (\ref{continuumQ}) on $\R^4$. Most importantly, it is easy to see that the flavor and baryon contribution  to the BCF anomaly is reproduced by the topological couplings in (\ref{r3kinetic}); the normalizations $\oint N_{f} A^{f(1)} \in 2 \pi \Z$, $\oint B^{f(2)} N_{f} \in 2 \pi \Z$, along with the corresponding ones for the $A^{c(1)}, B^{c(2)}$ fields given earlier, are important in showing this.

{\flushleft{\bf Nonperturbative potential and vacua:}}  After the somewhat lengthy exposition of symmetries and the matching of the BCF anomaly, we now come to the dynamics of the $\R^3 \times \S^1$ theory.  Potential terms in the $\R^3$ effective Lagrangian reflect both the $\S^1$-center stabilization and the semiclassical nonperturbative dynamics. The center stabilization is due to the   massive adjoint fermions added to stabilize the center. The field $\vec\phi$ is stabilized at the center symmetric value, $\vec\phi = {2 \pi \vec{\rho} \over N_c}$ ($\vec\rho$ is the Weyl vector, $\vec\rho = \sum_{k=1}^{N_c-1} \vec{w}_k$), acquires mass  $\sim {g \sqrt{ N_c}\over L}$, and does not participate in the low-energy dynamics \cite{Unsal:2008ch}. Thus,  we ignore $\vec\phi$ and consider   
the nonperturbative potential, which depends only on the lighter axion field and the dual photons.

 To leading exponential accuracy in the $v \gg {1\over N_c L} \gg \Lambda$ limit, the potential is given by the deformed Yang-Mills theory potential with a dynamical $\theta$-parameter. 
The nonperturbative potential,  generated by monopole-instantons, is
 \begin{eqnarray}
 \label{r3potential}
 V(\vec\sigma, a) = L^{-3} e^{- {8 \pi^2 \over g^2 N_c}} \sum\limits_{k=1}^{N_c} \left(1 - \cos\left(\vec\alpha_k \cdot \vec\sigma + {a N_f T_{\cal{R}} \over N_c}\right)\right)~,
  \end{eqnarray}
where $\vec\alpha_{k}$, $k=1,...N_c$, are the simple and affine roots of the $SU(N_c)$ algebra (the affine, or lowest, root is $\vec\alpha_{N_c} = - (\vec\alpha_1 + \ldots \vec\alpha_{N_c-1})$). The nonperturbative factor in front of the potential is associated with the monopole instanton action, $e^{-S_0} = e^{-{8 \pi^2 \over g^2 N_c}}$. The potential (\ref{r3potential}) is given up to an inessential multiplicative factor and a constant, ensuring $V \ge 0$, is added. From the discussion after (\ref{r3kinetic}), the $\Z_{N_f T_{\cal{R}}}$ shift symmetry $a \rightarrow a + \frac{2 \pi}{N_f T_{\cal{R}}}$ also  acts on $\vec\sigma$; in the center-stabilized vacuum $\vec\sigma \rightarrow \vec\sigma - {2 \pi \over N_c} \vec\rho$. We note that this symmetry action is that of an enhanced symmetry,\footnote{\label{enhancement}This symmetry enhancement is specific to  the calculable deformed-Yang-Mills theory setup that we consider here. Consider our $SU(N_f)$ theory with $N_f$ Dirac flavors in ${\cal{R}}$, with  $n_f$ Weyl flavors of  adjoint fields with a flavor-diagonal mass $m e^{i {\beta \over n_f}}$ added to ensure center stability. Here, we  think of $\beta$ as another nondynamical (``spurion") axion field (recall that $n_f \ge2$ with $m \sim 1/L$ is needed for center stability). A monopole-instanton vertex on $\R^3 \times \S^1$, in the limit $v \gg L$, i.e. with the $\cal{R}$ fermions  integrated out,  is $\sim e^{- {8 \pi^2 \over N_c g^2}} m^{n_f}  e^{i \vec\alpha_p \cdot \vec{\sigma}} e^{i \beta} e^{i  {a N_f T_{\cal{R}} \over N_c}}$, $p=1, \ldots N_c$. The $m^{n_f}$ factor is due to the lifting of zero-modes in the monopole-instanton background and the other factors follow from \cite{Unsal:2007jx}. This nonperturbative vertex is  invariant under an exact  $U(1)$ symmetry shifting both $\beta$ and $a$, as well as the $\Z_{2 N_c n_f}$ anomaly-free   discrete chiral symmetry of the adjoint fields, acting as $\beta  \rightarrow \beta + {2 \pi \over N_c}$ and $\vec\sigma \rightarrow \vec\sigma - {2 \pi \over N_c }\vec\rho$ (notice that these are the only exact chiral symmetries of the mixed adjoint-$\cal{R}$ theories coupled to axions). Giving an  expectation value to the spurion $\beta$ breaks the $U(1)$ but leaves the $\Z_{\cal{Q}}$, ${\cal{Q}} = {\rm lcm}(N_c, N_f T_{\cal{R}})$, symmetry, acting on $\vec\sigma$ by the above $2\pi\vec\rho/N_c$ shifts and on $a$ by $2\pi/(N_f T_{\cal{R}})$ shifts.  This is an emergent $0$-form symmetry  on $\R^3\times \S^1$. It acts on 't Hooft loop operators winding around the circle, $e^{i \vec\alpha_p \cdot \vec{\sigma}}$ \cite{Argyres:2012ka,Anber:2015wha}, and is not detectable by local physics on $\R^4$; for example, an instanton vertex on $\R^4$  does not exhibit an enhancement of the $\Z_{N_f T_{\cal{R}}}$ axion shift symmetry.} $\Z_{\cal{Q}}$, with ${\cal{Q}} \equiv {\rm lcm} (N_c, N_f T_{\cal{R}})$.

As already discussed, in the limit we study, the potential  (\ref{r3potential}) has a $0$-form emergent center symmetry,  in addition to the $\R^3$ $1$-form center. Its action is such that it cyclically permutes the various monopole-instanton factors, i.e. $\vec\alpha_k \cdot \vec\sigma \rightarrow \vec\alpha_{k+1({\rm mod} N_c)} \cdot \vec\sigma$. In group theoretic terms, this transformation is  $\vec\sigma \rightarrow {\cal P} \vec{\sigma}$, where ${\cal{P}}$ denotes an ordered product of Weyl reflections w.r.t.~all simple roots  \cite{Anber:2015wha}. This action of the $0$-form center symmetry transformation on the dual photons follows after restricting $\vec\phi$ to the Weyl chamber, see also \cite{Aitken:2017ayq}. For  use below,  we note the action of  ${\cal P}$ on the Weyl vector $\vec\rho$ and fundamental weights $\vec{w}_k$, $k=1,\ldots N_c -1$   \cite{Cox:2019aji}:
\begin{eqnarray} \label{zeroform2}
\Z_{N_c}^{(0)}: {\cal P}^l  \; {k  \over N_c} \vec\rho &=& {k\over N_c} \vec\rho - k \;\vec{w}_l~, ~l =1, \ldots N_c,~ {\rm with} ~\vec{w}_{N_c} \equiv 0.
\end{eqnarray}

To find the minima of the potential (\ref{r3potential}), we first extremize $V$ w.r.t. $\vec\sigma$. For any value of $a$, $V$ has $N_c$ extrema with respect to $\vec\sigma$ which lie in the unit cell of the weight lattice. These are at $\vec\sigma_q=  {2 \pi q\over N_c} \vec\rho$, $q=0,...N_c-1$. That these are extrema  can be easily verified, using $\vec\alpha_{k} \cdot \vec\rho = 1$, $k=1,...,N_c-1$ and $\vec\alpha_{N_c} \cdot \vec\rho = 1- N_c$ (it is  known \cite{Thomas:2011ee} that these are all extrema within the unit cell, see also \cite{Aitken:2018mbb}). 
Further, extremizing with respect to $a$,  we solve  ${d V\over da}(\vec\sigma_q, a) = N \sin\left( {2 \pi q + N_f T_{\cal{R}} a \over N_c}\right) =0$ to obtain $N_f T_{\cal{R}} a_{k,q} = \pi N_c k - 2 \pi q $, $k \in \Z$, $q =0,...N_c-1$. The potential (\ref{r3potential}) is nonnegative, hence the minima all have $V(\vec\sigma_q, a_{k,q})=0$, implying that $1  - \cos \pi k = 0$, so that $k=2 r$ is an even integer. Thus, the physically distinct minima of the potential for $a$ occur 
  at $a = {2 \pi (N_c r -  q) \over N_f T_{\cal{R}} }$, where $q$ takes the values from $0$ to $N_c-1$ and $r \in \Z$.
Counting the distinct ground states of (\ref{r3potential}) then reduces to finding pairs of integers  $q \in [0,\ldots, N_c -1]$ and $r$ such that 
the pairs of expectation values $(\langle e^{i a(x)}\rangle, \langle e^{i {\vec\alpha_k \cdot \vec\sigma(x)}}\rangle)= (e^{i  {2 \pi (N_c r - q) \over N_f T_{\cal{R}} }}, e^{i {2 \pi q \over N_c}})$ are distinct.
\begin{figure}[h]
   \centering
  \includegraphics[width=.60\linewidth]{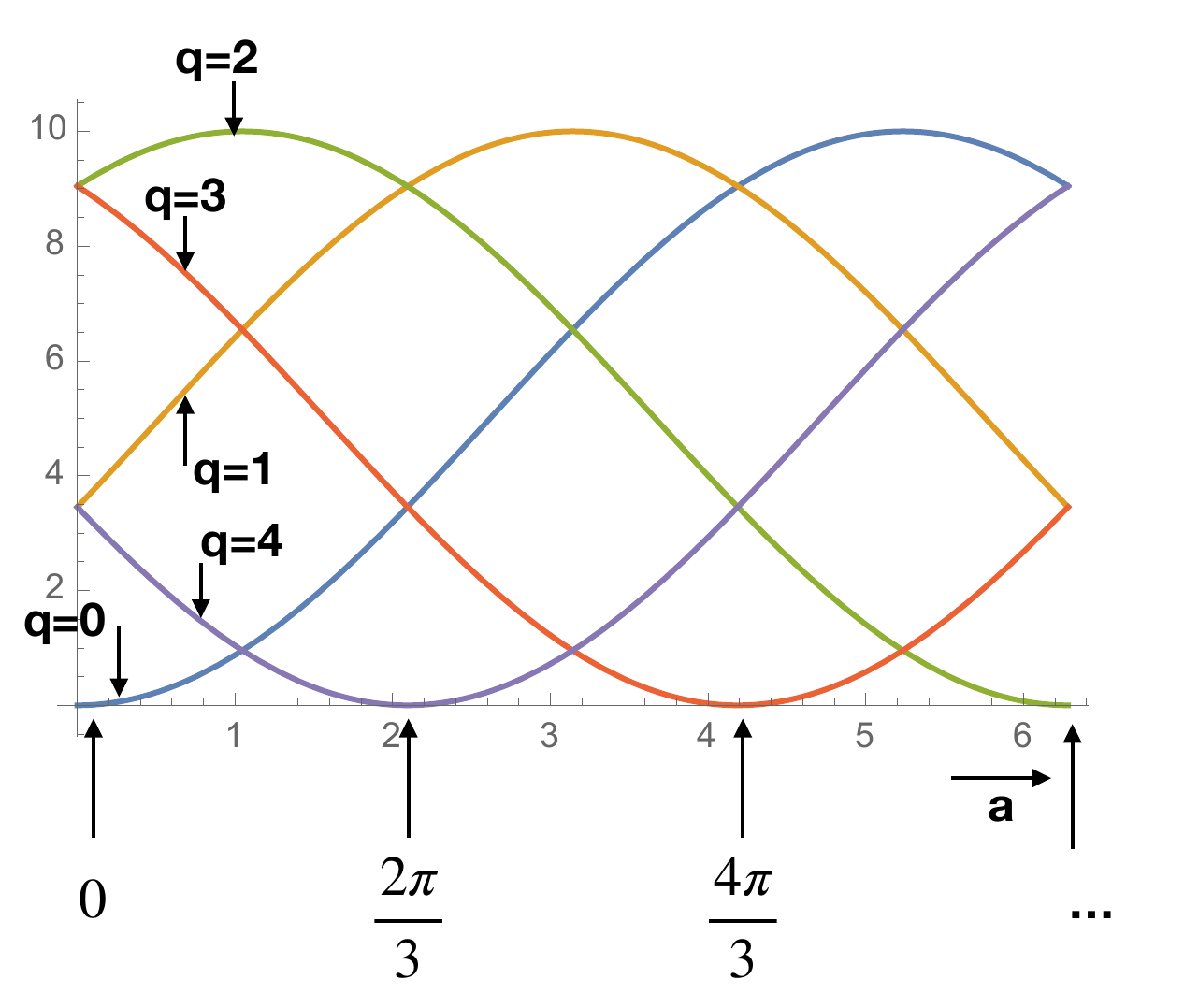}
  \caption{Axion potential  in $SU(5)$ QCD(F) with $N_f=3$: the five branches  $V(\vec\sigma_q, a)$, Eq.~(\ref{r3potential}) evaluated at $\vec\sigma_q = {2 \pi q \over 5} \vec\rho$, $q=0,1,2,3,4$. Shown are three of the $15$ $[{\cal{Q}}={\rm lcm}(5,3)]$ minima, determined as described in the text, related by the broken axion shift symmetry. It is clear that a change of the light field (axion) vacuum entails a rearrangement of the heavy confining degrees of freedom (the dual photons). In an axion  EFT this would be described by a transition from one branch of the axion potential to another one. This ``branch-hopping" on axion DW  is crucial in explaining the deconfinement of quarks.}  \label{fig:su5F}
\end{figure}
In all cases we have studied, there are ${\cal{Q}} \equiv {\rm lcm} (N_c, N_f T_{\cal{R}})$ values of $q$ and $r$ obeying the above conditions, corresponding to the spontaneous breaking of the $\Z_{N_f T_{\cal{R}}}$ (enhanced to $\Z_{\cal{Q}}$) axion shift  symmetry of interest to us.

For our discussion of deconfinement below, it is  important to note that the $0$-form symmetry $\Z_{N_c}^{(0)}$ is unbroken in the $N_f T_{\cal{R}}$ vacua of the axion theory: the transformation (\ref{zeroform2}) implies that the $\Z_{N_c}^{(0)}$ action maps $\vec\sigma_q$ to $\vec\sigma_q + 2 \pi \Z \vec{w}$, i.e.~to itself modulo weight-vector shifts, which are identifications of the compact dual photon fields (equivalently, the expectation values of the ``covering space" coordinates $e^{i \vec\alpha_k \cdot \vec\sigma}$, invariant under weight-vector shifts, are invariant under   $0$-form center transformations).  
 
The masses of the dual photons in the minima of (\ref{r3potential}) are given by the usual exponentially small nonperturbative scale, $m_\sigma^2 \sim L^{-2} {e^{- {8 \pi^2 \over g^2 N_c}}}$. The axion is significantly lighter due to the high axion scale, $m_a^2 \sim m_\sigma^2 \left({N_f T_{\cal{R}} \over N_c}\right)^2 {1 \over v^2 L^2}$. Thus, one would expect that the dual photons can be integrated out and the long distance dynamics and vacua be described in terms of a potential involving the axion field only. However,  as is clear from the discussion of the vacua, different vacua for the axion require a rearrangement of the heavy degrees of freedom, the dual photons. As we shall see, this is crucial to explaining deconfinement on DW. 

It is useful to plot the axion potential for one of the benchmark theories discussed in the paragraph after (\ref{globalchiral}). We consider QCD(F) with  $N_c = 5$, $N_f=3$ ($T_{\cal{R}}=1$). On Fig.~\ref{fig:su5F}, 
we show the function $V(\vec\sigma_q, a)$, i.e. the potential (\ref{r3potential}), evaluated at the $q$-th extremum w.r.t. $\vec\sigma$, as a function of the axion field $a$. Following the procedure for finding minima described earlier,  
we focus on the  three minima of the potential shown, all of  which  lie on different ``branches," i.e. have different values of $\vec\sigma_q$. Consider  the two neighboring minima at $a={2 \pi \over 3}$ and  $a = {4 \pi \over 3}$. They correspond to $\vec\sigma_q$ with $q=4$ and $q=3$, respectively. 
Thus, a DW interpolating between these two minima must necessarily  involve a change not only of $a$, but of  $\vec\sigma$ as well,  $\Delta \vec\sigma = {2 \pi \over 5}(4 - 3)\vec\rho = {2 \pi \over 5}\vec\rho$. 
This example shows that a change of the vacuum of the light degree of freedom (the axion) entails a rearrangement of the heavy degrees of freedom (the dual photons). We next explain, following \cite{Anber:2015kea}, how this rearrangement implies that quarks are deconfined on the axion DW.
   \begin{figure}[h]
    \centering
   \includegraphics[width=0.9\linewidth]{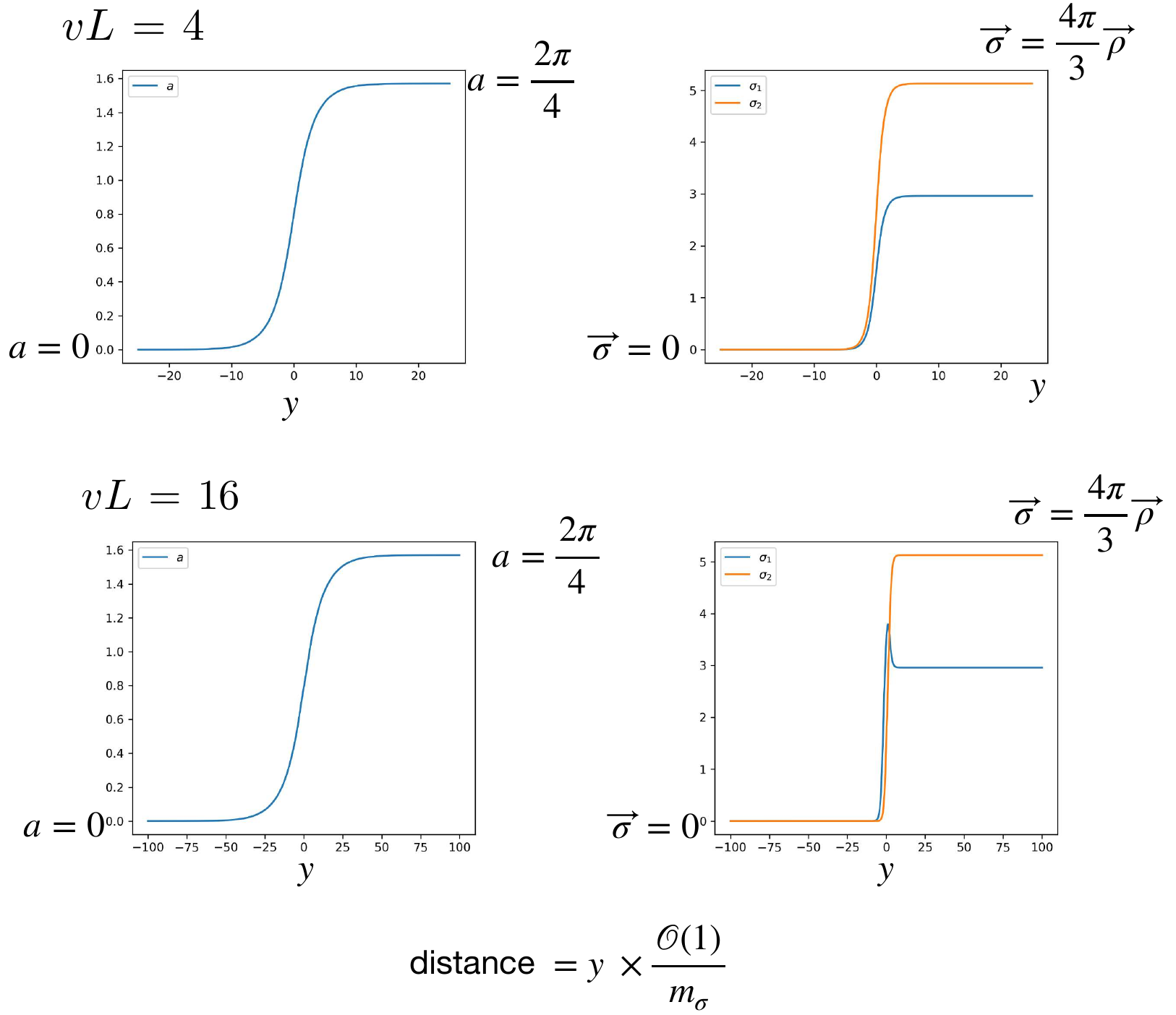}
   \caption{The axion DW between two neighboring vacua $(a=0, \vec\sigma=0)$ and $(a = {2 \pi \over 4}, \vec\sigma = {4 \pi \over 3} \vec\rho)$ for $SU(3)$ QCD(F) with $N_f=4, N_f T_{\cal{R}}=4$. On the top two panels, we show the wall profiles for $a$ and $\vec\sigma$ for $v L = 4$ and on the lower two panels, for $vL = 16$. Distances are measured in terms of the inverse dual photon wavelength, the ``confinement scale". The multi-scale structure of the DW is evident in the much longer scale of variation of the axion profile for larger $vL$. The overall width of the wall is set by the axion Compton wavelength, while the electric flux---the region with nonzero gradient of $\vec\sigma$---carried by the DW is squeezed into  a much narrower region, of order the confinement scale.  (This figure was provided to us by Andrew Cox and Samuel Wong. The methods used to obtain the figure are described in \cite{Cox:2019aji}.) }\label{fig:axiondw}
 \end{figure}
 {\flushleft{\bf Domain walls and deconfinement:}} A consideration of the mass scales leads one to expect that the main contribution to the DW dynamics is given by the axion field and  that its Compton wavelength alone determines the DW properties.We notice that this feature is shared by the naive  axion effective theory (\ref{effective2}) on $\R^4$ where the color degrees of freedom are integrated out. However, it is also clear that if the DW on $\R^4$ are supposed to deconfine quarks, the long-distance axion approximation cannot be the whole story, as it knows nothing about color flux and confinement.

 The expectation of a multi-scale structure of the axion DW can be clearly seen in our calculable $\R^3 \times \S^1$ set up. The axion DW are   more complex objects and also involve scales set by $m_\sigma$, the nonperturbative scale determining the mass gap for gauge fluctuations---this nonperturbative scale plays the role of $\Lambda$ in the $\R^4$ theory. A numerical solution for a DW profile showing the multiscale structure is  on Fig.~\ref{fig:axiondw}. This multi-scale structure of DW is related to the quark deconfinement on DW. It is natural to expect that different scales associated with the axion DW are also important to the physics of  quark  deconfinement on axion DW on $\R^4$.

  \begin{figure}[h]
    \centering
   \includegraphics[width=.80\linewidth]{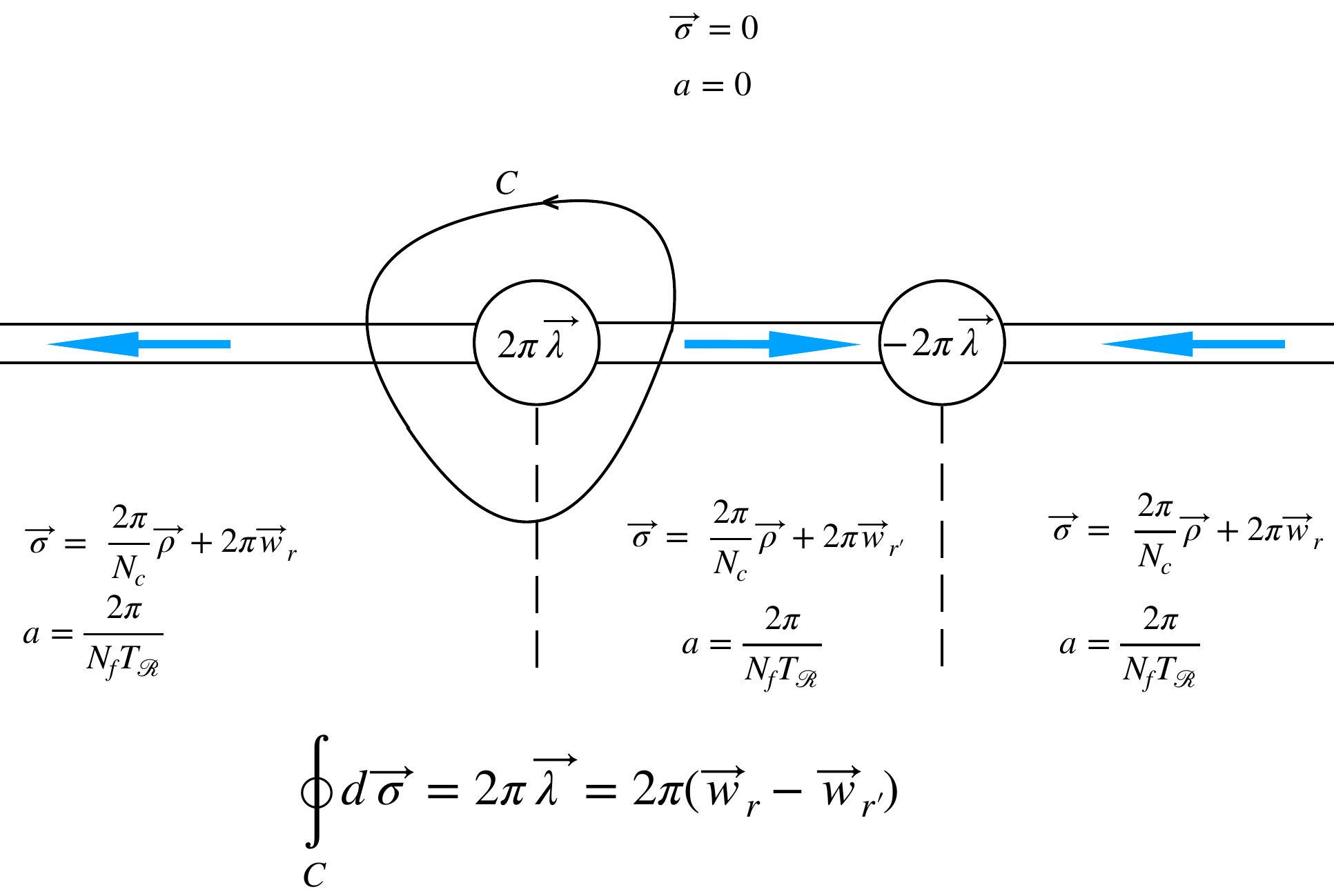}
   \caption{The mechanism of quark deconfinement on axion DW. The quark/antiquark pair shown is suspended on a DW between two vacua of the axion theory (\ref{r3potential}).  
     The monodromy of the dual photon $\oint_C d \vec\sigma$ around the junction of two degenerate (due to the $0$-form $\Z_{N_c}^{(0)}$ center symmetry) DW  equals the electric charge of the quark, or its weight   $2 \pi \vec\lambda$ (the dashed lines are where weight-lattice discontinuities of $\vec\sigma$ occur, but no physical discontinuity). The  electric fluxes of the quark and antiquark are absorbed by the DW, as indicated by the arrows. The quarks experience no force, due to the equal tensions of the DW to the left and right of each quark. See \cite{Cox:2019aji}  for detailed explanations of a similar mechanism in super-Yang-Mills theory and for results of actual numerical simulations of DW with suspended quark-antiquark pairs.}\label{fig:deconfwall}
 \end{figure}
 
 To see how deconfinement works in our calculable setup (illustrated on Fig.~\ref{fig:deconfwall})
recall  that the monodromy (i.e.~change) of $\vec\sigma$ across the DW is equal to the electric flux carried along the  DW worldvolume, as implied by the duality   $\partial_x \vec{\sigma} \sim   \vec{E}_y, \partial_y \vec{\sigma} \sim   -\vec{E}_x$ ($x,y$ are the spatial coordinates). Further, we recall that Cartan subalgebra electric fluxes of quarks of  $N$-ality $k$ are   $2 \pi \vec{w}_k + 2 \pi \times ({\rm root \;vectors})$. On the other hand, as we saw above, the fluxes carried by axion  DW are ${2 \pi \over N_c}\vec{\rho}$ (for simplicity,  we focus on DW where $\vec\sigma$ jumps from one branch to the neighboring one, i.e. $|\Delta q| = 1$, as for neighboring vacua on Fig.~\ref{fig:su5F}; the discussion for non-neighboring DW proceeds similarly).

The ${2 \pi \over N_c}\vec{\rho}$ electric flux carried by the DW is, however, only a fraction of the flux carried by quarks. However, we now recall that our theory has a $\Z_{N_c}^{(0)}$ center symmetry (\ref{zeroform2}), unbroken in the $N_f T_{\cal{R}}$ vacua. This symmetry maps DW solutions to DW solutions, and implies that   there are $N_c$ DW solutions, with equal tensions, between any of the two axion vacua. However, these walls carry different electric fluxes:  the first relation in (\ref{zeroform2}), with $k=1$, implies that the $N_c$ different domain walls of the same tension carry electric fluxes 
${2 \pi \over N_c} \vec\rho + 2 \pi \vec{w}_r$, with $r=0,...N_c-1$ (recall that $\vec w_0 \equiv 0$). Thus the difference between the electric fluxes carried by two walls, labelled by $r, r'$, is $2 \pi (\vec{w}_r - \vec{w}_{r'})$.  Thus, as illustrated on Figure~\ref{fig:deconfwall}, the junction of two such DW supports quarks of weights $2 \pi \vec\lambda = 2 \pi (\vec{w}_r - \vec{w}_{r'})$. A quark/antiquark pair suspended on two consecutive junctions experiences no force, due to the equal tensions of the DW. It is clear that any weight of the fundamental representation is deconfined on this $\Delta q=1$ wall. Similar to  the 
 discussion of \cite{Cox:2019aji}, this result for the weights of  quarks suspended on $|\Delta q| = 1$ DW also implies that Wilson loops for any nonzero $N$-ality representation  also show  perimeter law on the wall.

{\bf {\flushleft{Acknowledgments:}}} MA acknowledges the hospitality at the University of Toronto, where this work was completed. MA is supported by the NSF grant PHY-1720135.  EP is supported by a NSERC Discovery Grant. We are grateful to Andrew Cox and  Samuel Wong for their numerical solution of the multi-scale axion DW and to Michael Luke for discussions.

  \bibliography{ReferencesADW.bib}
  
  \bibliographystyle{JHEP}
  \end{document}